\documentclass[runningheads]{llncs}
\usepackage{graphicx}
\graphicspath{ {./img/} }
\usepackage{tabularx}
\usepackage{amsfonts}
\usepackage{amsmath}
\usepackage{subcaption}
\usepackage[hidelinks]{hyperref}

\newcolumntype{Y}{>{\centering\arraybackslash}X}
\begin{document}
\title{Squeeze-and-Excitation Normalization for Automated Delineation of Head and Neck Primary Tumors in Combined PET and CT Images\thanks{\textbf{Cite this paper as:} Iantsen A., Visvikis D., Hatt M. (2021) Squeeze-and-Excitation Normalization for Automated Delineation of Head and Neck Primary Tumors in Combined PET and CT Images. In: Andrearczyk V., Oreiller V., Depeursinge A. (eds) Head and Neck Tumor Segmentation. HECKTOR 2020. Lecture Notes in Computer Science, vol 12603. Springer, Cham. \url{https://doi.org/10.1007/978-3-030-67194-5_4}}}
\titlerunning{SE Normalization for Delineation of H\&N Tumors in PET/CT}
\author{Andrei Iantsen\orcidID{0000-0002-6690-6070} \and
Dimitris Visvikis\orcidID{0000-0003-0831-3637} \and
Mathieu Hatt\orcidID{0000-0002-8938-8667}}
\authorrunning{A. Iantsen et al.}
\institute{LaTIM, INSERM, UMR 1101, University Brest, Brest, France}
\maketitle              % typeset the header of the contribution
\begin{abstract}
Development of robust and accurate fully automated methods for medical image segmentation is crucial in clinical practice and radiomics studies. In this work, we contributed an automated approach for Head and Neck (H\&N) primary tumor segmentation in combined positron emission tomography / computed tomography (PET/CT) images in the context of the MICCAI 2020 Head and Neck Tumor segmentation challenge (HECKTOR). Our model was designed on the U-Net architecture with residual layers and supplemented with Squeeze-and-Excitation Normalization. The described method achieved competitive results in cross-validation (DSC 0.745, precision 0.760, recall 0.789) performed on different centers, as well as on the test set (DSC 0.759, precision 0.833, recall 0.740) that allowed us to win first prize in the HECKTOR challenge among 21 participating teams. The full implementation based on PyTorch and the trained models are available at \url{https://github.com/iantsen/hecktor}. 

\end{abstract}
\section{Introduction}
Combined positron emission tomography / computed tomography (PET/CT) imaging is broadly used in clinical practice for radiotherapy treatment planning, initial staging and response assessment. In radiomics analyses, quantitative evaluation of radiotracer uptake in PET images and tissues density in CT images, aims at extracting clinically relevant features and building diagnostic, prognostic and predictive models. The segmentation step of the radiomics workflow is the most time-consuming bottleneck and variability in usual semi-automatic segmentation methods can significantly affect the extracted features, especially in case of manual segmentation, which is affected by the highest magnitude of inter- and intra-observer variability. Under these circumstances, a fully automated segmentation is highly desirable to automate the whole process and facilitate its clinical routine usage.

The MICCAI 2020 Head and Neck Tumor segmentation challenge (HECKTOR)~\cite{ref_article1} aims at evaluating automatic algorithms for segmentation of Head and Neck (H\&N) tumors in combined PET and CT images. A dataset of 201 patients from four medical centers in Qu\'{e}bec (CHGJ, CHMR, CHUM and CHUS) with histologically proven H\&N cancer in the oropharynx is provided for a model development. A test set comprised of 53 patients from a different center in Switzerland (CHUV) is used for evaluation. All images were re-annotated by an expert for the purpose of the challenge in order to determine primary gross tumor volumes (GTV) on which the methods are evaluated using the Dice score (DSC), precision and recall.

This paper describes our approach based on convolutional neural networks supplemented with Squeeze-and-Excitation Normalization (SE Normalization or SE Norm) layers to address the goal of the HECKTOR challenge.

\section{Materials \& Methods}
\subsection{SE Normalization}
The key element of our model is SE Normalization layers~\cite{ref_article6} that we recently proposed in the context of the Brain Tumor Segmentation Challenge (BraTS 2020)~\cite{ref_article3}. Similarly to Instance Normalization~\cite{ref_article4}, for an input ${\mathrm{X} = (x_{1}, x_{2}, \dots, x_{\mathrm{N}})}$ with $\mathrm{N}$ channels, SE Norm layer first normalizes all channels of each example in a batch using the mean and standard deviation:

\begin{equation}
x'_{i} = \frac{1}{\sigma_{i}}(x_{i} - \mu_{i})
\end{equation}

where $\mu_{i} = \textrm{E}[x_{i}]$ and $\sigma_{i} = \sqrt{\textrm{Var}[x_{i}] + \epsilon}$ with $\epsilon$ as a small constant to prevent division by zero. After, a pair of parameters $\gamma_{i}, \beta_{i}$ are applied to each channel to scale and shift the normalized values:

\begin{equation}
y_{i} = \gamma_{i}x'_{i} + \beta_{i}
\end{equation}

\begin{figure}
	\centering
	\begin{subfigure}{0.4\textwidth}
		\centering
		\includegraphics[scale=1.25]{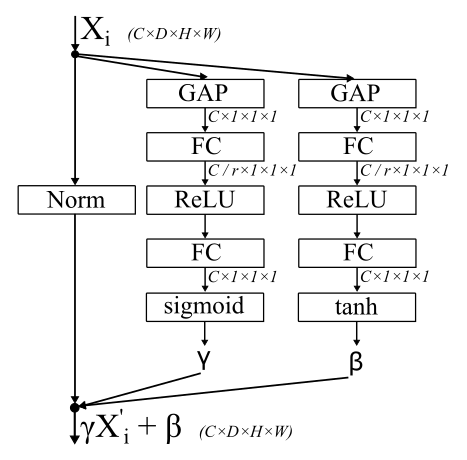}
		\phantomsubcaption\label{senorm} %{SE Normalization layer}
	\end{subfigure}%
	\begin{subfigure}{0.3\textwidth}
		\centering
		\includegraphics[scale=1.4]{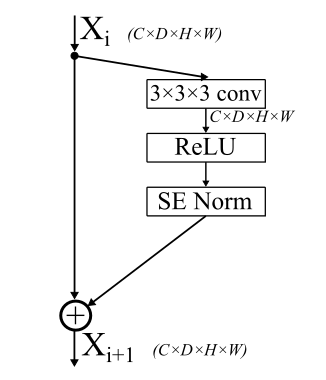}
		\phantomsubcaption\label{resblock1} % {Residual layer with the shortcut connection}
	\end{subfigure}%
	\begin{subfigure}{0.3\textwidth}
		\centering
		\includegraphics[scale=1.4]{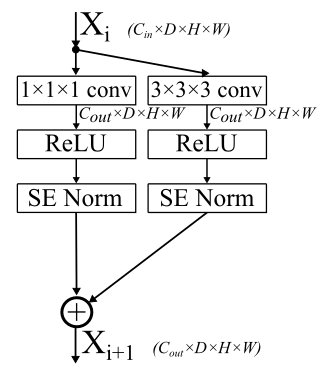}
		\phantomsubcaption\label{resblock2} % {Residual layer with the non-linear projection}
	\end{subfigure}%
	\caption{Layers with SE Normalization: (\subref{senorm}) SE Norm layer, (\subref{resblock1}) residual layer with the shortcut connection, and (\subref{resblock2}) residual layer with the non-linear projection. Output dimensions are depicted in italics.}\label{fig1}
\end{figure}

In case of Instance Normalization, both parameters $\gamma_{i}, \beta_{i}$, fitted in the course of training, stay fixed and independent on the input $\mathrm{X}$ during inference. By contrast, we propose to model the parameters $\gamma_{i}, \beta_{i}$ as functions of the input $\mathrm{X}$ by means of Squeeze-and-Excitation (SE) blocks~\cite{ref_article5}, i.e

\begin{align} 
	\gamma &= f_{\gamma}(X) \\ 
	\beta &= f_{\beta}(X)
\end{align}

where ${\gamma = (\gamma_{1}, \gamma_{2}, \dots, \gamma_{\mathrm{N}})}$ and ${\beta = (\beta_{1}, \beta_{2}, \dots, \beta_{\mathrm{N}})}$ - the scale and shift parameters for all channels, $f_{\gamma}$ - the original SE block with the sigmoid, and $f_{\beta}$ is modeled as the SE block with the tanh activation function to enable the negative shift (see Fig.~\ref{senorm}). Both of SE blocks first apply global average pooling (GAP) to squeeze each channel into a single descriptor. Then, two fully connected (FC) layers aim at capturing non-linear cross-channel dependencies. The first FC layer is implemented with the reduction ratio $r$ to form a bottleneck for controlling model complexity. Throughout this paper, we apply SE Norm layers with the fixed reduction ration $r = 2$.

\subsection{Network Architecture}
Our model is built upon a seminal U-Net architecture~\cite{ref_article7,ref_article8} with the use of SE Norm layers~\cite{ref_article6}. Convolutional blocks, that form the model decoder, are stacks of $3\times3\times3$ convolutions and ReLU activations followed by SE Norm layers. Residual blocks in the encoder consist of convolutional blocks with shortcut connections (see Fig.~\ref{resblock1}). If the number of input/output channels in a residual block is different, a non-linear projection is performed by adding the $1\times1\times1$ convolutional block to the shortcut in order to match the dimensions (see Fig.~\ref{resblock2}).  

\begin{figure}
	\includegraphics[width=\textwidth]{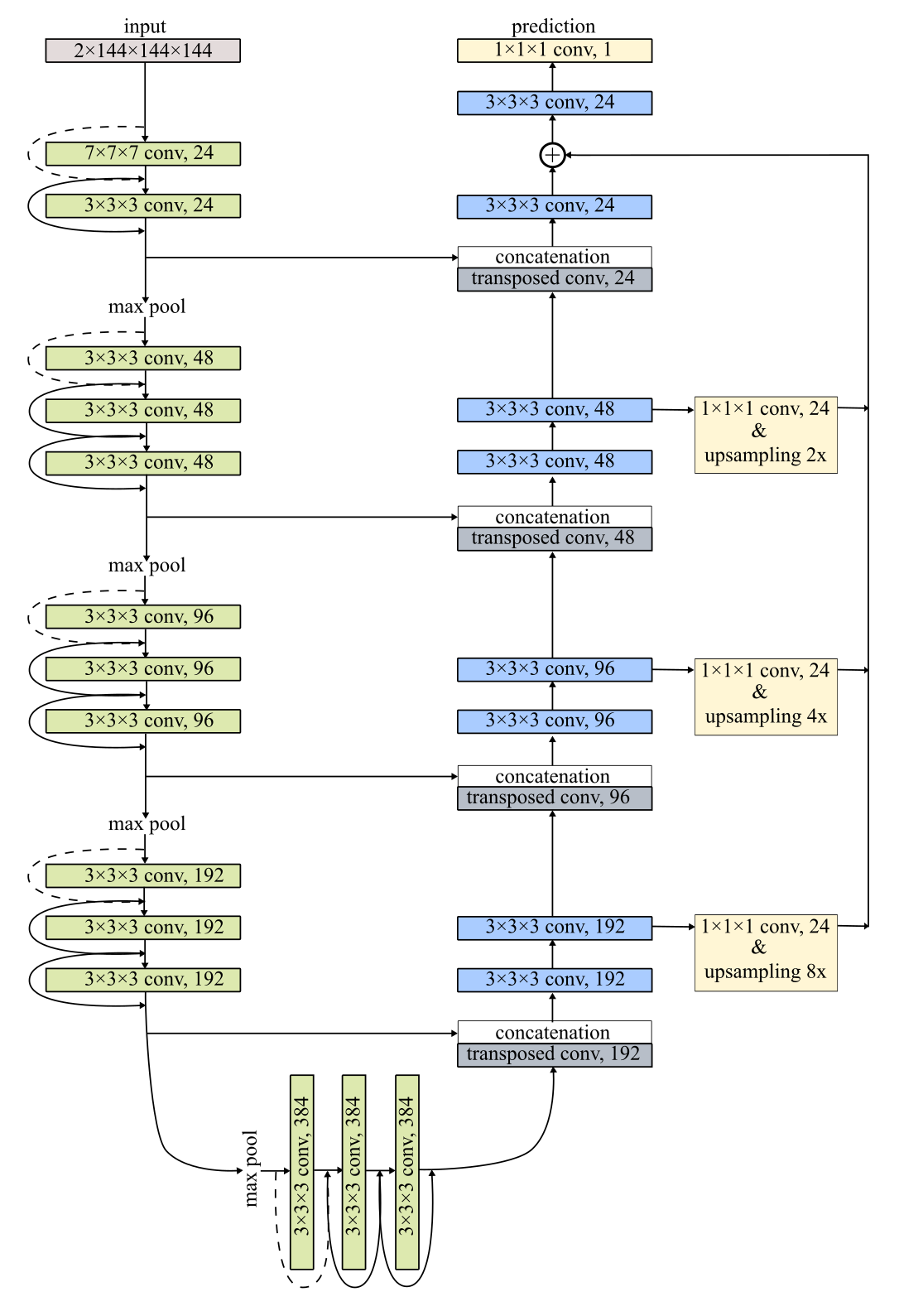}
	\caption{The model architecture with SE Norm layers. The input consists of PET/CT patches of the size of ${144\times144\times144}$ voxels. The encoder consists of residual blocks with identity (solid arrows) and projection (dashed arrows) shortcuts. The decoder is formed by convolutional blocks. Additional upsampling paths are added to transfer low-resolution features further in the decoder. Kernel sizes and numbers of output channels are depicted in each block.} \label{fig2}
\end{figure}

In the encoder, downsampling is done by applying max pooling with the kernel size of $2\times2\times2$. To linearly upsample feature maps in the decoder, $3\times3\times3$ transposed convolutions are used. In addition, we supplement the decoder with three upsampling paths to transfer low-resolution features further in the model by applying the $1\times1\times1$ convolutional block to reduce the number of channels, and utilizing trilinear interpolation to increase the spatial size of the feature maps (see Fig.~\ref{fig2}, yellow blocks).

The first residual block placed after the input is implemented with the kernel size of $7\times7\times7$ to increase the receptive field of the model without significant computational overhead. The sigmoid function is applied to output probabilities for the target class.

\subsection{Data Preprocessing \& Sampling}
Both PET and CT images were first resampled to a common resolution of ${1\times1\times1}$ $mm^3$ with trilinear interpolation. Each training example was a patch of ${144\times144\times144}$ voxels randomly extracted from a whole PET/CT image, whereas validation examples were received from bounding boxes provided by organizers. Training patches were extracted to include the tumor class with the probability of 0.9 to facilitate model training.

CT intensities were clipped in the range of ${\left[ -1024, 1024\right]}$ Hounsfield Units and then mapped to ${\left[ -1, 1\right]}$. PET images were transformed independently with the use of Z-score normalization, performed on each patch.

\subsection{Training Procedure}
The model was trained for 800 epochs using Adam optimizer on two GPUs NVIDIA GeForce GTX 1080 Ti (11 GB) with a batch size of 2 (one sample per worker). The cosine annealing schedule was applied to reduce the learning rate from $10^{-3}$ to $10^{-6}$ within every 25 epochs.

\subsection{Loss Function}
The unweighted sum of the Soft Dice Loss~\cite{ref_article9} and the Focal Loss~\cite{ref_article10} is utilized to train the model. Based on~\cite{ref_article9}, the Soft Dice Loss for one training example can be written as

\begin{equation}
L_{Dice}(y, \hat{y}) = 1 - \frac{2\sum_{i}^{\mathrm{N}} y_{i} \hat{y}_{i} + 1} 
{\sum_{i}^{\mathrm{N}} y_{i} + \sum_{i}^{\mathrm{N}} \hat{y}_{i} + 1}
\end{equation}

The Focal Loss is defined as

\begin{equation}
L_{Focal}(y, \hat{y}) = - \frac{1}{\mathrm{N}} \sum_{i}^{\mathrm{N}} y_{i}(1 - \hat{y}_{i})^{\gamma}\ln(\hat{y}_{i})
\end{equation}

In both definitions, ${y_{i} \in \{ 0, 1 \} }$ - the label for the \textit{i-th} voxel, ${\hat{y}_{i} \in \left[ 0, 1 \right] }$ - the predicted probability for the \textit{i-th} voxel, and $\mathrm{N}$ - the total numbers of voxels. Additionally we add +1 to the numerator and denominator in the Soft Dice Loss to avoid the zero division in cases when the tumor class is not present in training patches. The parameter $\gamma$ in the Focal Loss is set at 2.

\begin{table}
	\centering
	\caption{The performance results on different cross-validation splits. Average results (the row 'Average') are provided for each evaluation metric across all centers in the leave-one-center-out cross-validation (first four rows). The mean and standard deviation of each metric are computed across all data samples in the corresponding validation center. The row 'Average (rs)' indicates the average results on the four random data splits.}\label{tab1}
	\begin{tabularx}{\textwidth}{Y|Y|Y|Y|}
		\hline
		Center & 
		DSC &
		Precision &
		Recall \\		
		\hline\hline
		
		CHUS (${n = 72}$) & 
		$0.744\pm0.206$ & $0.763\pm0.248$ &	$0.788\pm0.226$\\
		\hline
		
		CHUM (${n = 56}$) & 
		$0.739\pm0.190$ & $0.748\pm0.224$ &	$0.819\pm0.216$ \\
		\hline
		
		CHGJ (${n = 55}$) & 
		$0.801\pm0.180$ & $0.791\pm0.208$ &	$0.839\pm0.200$ \\
		\hline
		
		CHMR (${n = 18}$) & 
		$0.696\pm0.232$ & $0.739\pm0.286$ & $0.712\pm0.228$ \\
		\hline\hline
		
		Average & 
		0.745 & 0.760 & 0.789 \\
		\hline
		
		Average (rs) & 
		0.757 & 0.762 & 0.820 \\
		\hline
	\end{tabularx}
\end{table}

\subsection{Ensembling}
Our results on the test set were produced with the use of an ensemble of eight models trained and validated on different splits of the training set. Four models were built using a leave-one-center-out cross-validation, i.e, the data from three centers was used for training and the data from the fourth center was held out for validation. Four other models were fitted on random training / validation splits of the whole dataset. Predictions on the test set were produced by averaging predictions of the individual models and applying a threshold operation with a value equal to ${0.5}$.

\begin{table}
	\centering
	\caption{The test set results of the ensemble of eight models.}\label{tab2}
	\begin{tabularx}{\textwidth}{Y|Y|Y|Y|}
		\hline
		Center & 
		DSC &
		Precision &
		Recall \\		
		\hline\hline
		
		CHUV (${n = 53}$) & 
		0.759 & 0.833 & 0.740  \\
		\hline
	\end{tabularx}
\end{table}

\section{Results \& Discussion}

Our validation results in the context of the HECKTOR challenge are summarized in Table~\ref{tab1}. The best outcome in terms of all evaluation metrics was received for the 'CHGJ' center with 55 patients. The model demonstrated the poorest performance for the 'CHMR' center that is least represented in the whole dataset. The differences with the two other centers was minor for all evaluation metrics. The small spread between all centers and the average results implies that the model predictions were robust and any center-specific data standardization was not required. This finding is supported by the lack of significant difference in the average results between the leave-one-center-out and random split cross-validations.

The ensemble results on the test set consisting of 53 patients from the 'CHUV' center are presented in Table~\ref{tab2}. On the previously unseen data, the ensemble of eight models achieved the highest results among 21 participating teams with the Dice score of 75.9\%, precision 83.3\% and recall 74\%.


\end{document}